\documentclass[apj, iop, twocolumn, tighten, twocolappendix, numberedappendix]{aastex63}

\usepackage{times}
\usepackage{amsmath,amssymb,amsfonts}
\usepackage{bm}
\usepackage{graphicx}
\DeclareGraphicsExtensions{.pdf,.eps,.png,.jpg}
\usepackage{acronym}
\usepackage{color, colortbl}
\usepackage{xcolor}
\usepackage{subfigure}
\usepackage{tabularx}
\usepackage{array}
\usepackage{booktabs}
\usepackage{longtable}
\usepackage{multirow}
\usepackage{dcolumn}
\usepackage{soul}
\usepackage{float}
\usepackage{hyperref}
\usepackage{cleveref}
\usepackage{enumitem}

\allowdisplaybreaks

\newcommand{\msun}{\mbox{$M_\odot$}}

\newcolumntype{C}[1]{>{\centering\arraybackslash}p{#1}}
\newcommand{\bL}{\mbox{\emph{\textbf{L}}}}
\newcommand{\NA}{\mbox{\hspace{-.08em}$\cdot\hspace{-.17em}\cdot\hspace{-0.17em}\cdot$}}
\newcommand{\colsep}{-1.7mm}
\mathchardef\mhyphen="2D

\shortauthors{Kim et al.}
\begin{document}

\title{Deep Learning-based Search for Microlensing Signature from Binary Black Hole Events in GWTC-1 and -2}

\author[0000-0003-1653-3795]{Kyungmin~\surname{Kim}}
\email{kyungmin.kim@ewha.ac.kr}
\affiliation{Department of Physics, Ewha Womans University, 52 Ewhayeodae-gil, Seodaemun-gu, Seoul 03760, South Korea}
\affiliation{Korea Astronomy and Space Science Institute, 776 Daedeokdae-ro, Yuseong-gu, Daejeon 34055, South Korea}

\author[0000-0003-2510-4132]{Joongoo~\surname{Lee}}
\affiliation{Department of Physics and Astronomy, Seoul National University, Seoul 08826, South Korea}

\author[0000-0002-3887-7137]{Otto~A.~\surname{Hannuksela}}
\affiliation{Department of Physics, The Chinese University of Hong Kong, Shatin, New Territories, Hong Kong}

\author[0000-0003-4297-7365]{Tjonnie~G.~F.~\surname{Li}}
\affiliation{Department of Physics, The Chinese University of Hong Kong, Shatin, New Territories, Hong Kong}
\affiliation{Institute for Theoretical Physics, KU Leuven, Celestijnenlaan 200D, B-3001 Leuven, Belgium}
\affiliation{Department of Electrical Engineering (ESAT), KU Leuven, Kasteelpark Arenberg 10, B-3001 Leuven, Belgium}

\begin{abstract}
We present the result of the first deep learning-based search for the signature of microlensing in gravitational waves. This search seeks the signature induced by lenses with masses between $10^3$--$10^5$\msun~from spectrograms of the binary black hole events in the first and second gravitational-wave transient catalogs. We use a deep learning model trained with spectrograms of simulated noisy gravitational-wave signals to classify the events into two classes, lensed or unlensed. We introduce ensemble learning and a majority voting-based consistency test for the predictions of ensemble learners. The classification scheme of this search primarily classifies one event, GW190707\_093326, into the lensed class. To verify the primary classification of this event, we also examine the median probability to the lensed class and observe the resulting value, $0.984^{+0.012}_{-0.342}$, agrees with an empirical criterion $>\!0.6$ for claiming the detection of a lensed signal. However, the uncertainty of the estimated $p$-value for the median probability and error, ranging from 0 to 0.1, convinces us GW190707\_093326 is less likely a lensed event because it includes $p\!\geq\!0.05$ where the unlensed hypothesis is true. Therefore, we conclude our search finds no significant evidence of microlensing signature from the evaluated binary black hole events.
\end{abstract}




\section{Introduction}

To date, about 100 gravitational-wave (GW) events have been identified from the data taken by the Advanced LIGO~\citep{TheLIGOScientific:2014jea} and the Advanced Virgo~\citep{TheVirgo:2014hva} during the first, second, and third observing runs, referred to O1, O2, and O3, respectively.~\citep{LIGOScientific:2018mvr, LIGOScientific:2020ibl, LIGOScientific:2021usb, LIGOScientific:2021djp, Nitz:2018imz, Nitz:2020oeq, Nitz:2021uxj, Nitz:2021zwj, Olsen:2022pin} The progenitors of all events are recognized as distant compact binary mergers such as binary black holes, binary neutron stars, and neutron star-black hole binaries; it turned out that binary black hole (BBH) events get the majority in the population among the three source types. 

Thanks to the observation of GWs, the GW astronomy era has emerged. The advent of the GW astronomy has opened a new window for looking at the Universe, i.e., it enables us to tackle diverse phenomena in the Universe like electromagnetic waves-based astronomy (EM astronomy) has done for centuries. For example, we could understand more about the equation-of-state of the interior matter of neutron stars by observing GW170817~\citep{LIGOScientific:2018cki,LIGOScientific:2019eut} and could measure the Hubble parameter~\citep{LIGOScientific:2017adf} by jointly observing GW170817 and GRB170817A~\citep{LIGOScientific:2017ync}. Besides the existing scientific investigations, many more GW avenues still await us. In this context, one such avenue with a rich EM astronomical history is the gravitational lensing of GWs (or, simply, GW lensing).

In practice, many searches have tried to look for the signature
of GW lensing by now.~\citep{Hannuksela:2019kle, LIGOScientific:2021izm, Broadhurst:2019ijv,McIsaac:2019use,Pang:2020qow,Broadhurst:2020cvm,Dai:2020tpj,Liu:2020par,Diego:2021fyd} Specifically, \cite{Hannuksela:2019kle, LIGOScientific:2021izm} have conducted comprehensive searches exploring the signature of strong, weak, and microlensing in the forty-six BBH events reported in the first and second gravitational-wave transient catalogs referred to GWTC-1 and -2, respectively; there has been no widely accepted detection thus far.

Despite this, the forecasts on the detection rate make us expect detection of lensed GWs will be achievable in future observing runs: For example, \cite{KenKaze:2018prd} estimated $\sim \mathcal{O}(1)$ events per year for the strongly lensed event with the ground-based detectors' the design sensitivities reaching to the redshift $z\!\sim\!1$; for microlensing events, \cite{Diego:2019lcd} estimated similar rates under certain circumstances, e.g., the source is in the redshift interval $2\!<\!z\!<\!3$, and the magnification factor is $\sim\!30$. 

Meanwhile, machine learning (or deep learning)-based search methods, e.g.,~\cite{Goyal:2021hxv} for strongly lensed events and \cite{Kim:2020xkm} for microlensed events, also have been suggested. Particularly, in~\cite{Kim:2020xkm}, we established a novel method utilizing deep learning for identifying the signature of the microlensing in GWs (GW microlensing, hereafter). It has been discussed that GW microlensing can be recognized by its characteristic signature, \emph{beating patterns}, caused by the superposition of multiple lensed GW signals arriving to the GW detector network with about $\mathcal{O}(\textrm{msec})$ of time delays to each other.~\citep{Cao:2014oaa, Jung:2017flg, Christian:2018vsi, Diego:2019lcd, Diego:2019rzc, Pagano:2020rwj, LIGOScientific:2021izm, Seo_microlensing, Meena:2022unp} However, as discussed in the literature, the complex lens configuration embedded around macro lenses like galaxies or galaxy clusters, or relatively weaker signature than strong lensing make the search for GW microlensing become challenging. 

Instead, in the previous work, we assumed that the strong lensing occurred by lenses with masses between $10^3$\msun--$10^5$\msun~might induce short time delays---comparable to that of GW microlensing---between two lensed signals: This assumption alternatively makes beating patterns be shown on the waveform of BBH events as if the GW microlensing does because the desired time delay is much shorter than the typical duration of the observed BBH signals, $\lesssim\!1~\textrm{sec}$. We designed the method to seek such signature from spectrogram images of BBH events to bring the excellence of a state-of-the-art deep learning model, VGG-19~\citep{Simonyan14c}. We supposed an arbitrary detector is optimally positioned to the source BBH's orientation for simplicity of the proof-of-principle study. From the performance tests on two tasks, the classification of simulated signals and the regression of intrinsic and extrinsic parameters of those signals, we concluded that the method is feasible for identifying beating patterns from spectrograms of BBH events. 

In this search, we revisit the BBH events examined in~\cite{Hannuksela:2019kle} and \cite{LIGOScientific:2021izm} with the deep learning-based classification strategy built in~\cite{Kim:2020xkm}. Instead, we update the method to reflect reality a bit more, e.g., supposing the detection of such events is done via the GW detector network. Furthermore, we use the design sensitivity of the Advanced LIGO~\citep{Harry:2010zz} not only to mimic the irremovable noise presented in spectrograms of the BBH events but to regard a general noise model commonly applicable for the events observed by the different detectors operated with non-identical sensitivities over the three observing runs. On top of that, we introduce an ensemble learning to mitigate improperly biased prediction that might be occurred by using a single learner. For the initial classes on each detector's data, predicted by the ensemble learners, we employ a majority voting-based consistency test to classify the evaluated BBH events into two primary classes, lensed or unlensed.

We figure out that one event, GW190707\_093326, out of forty-six events is primarily classified as a lensed signal. To verify the result, we further investigate this event via the following tests: First, we examine the median probability to the lensed class of the event and observe that the result, $0.984^{+0.012}_{-0.342}$, agrees to an empirical criterion $>\!0.6$ for claiming the detection of a lensed signal. Second, from the uncertainty of estimated $p$-value for the median probability and error, ranging from 0 to 0.1, we, however, convince GW190707\_093326 is less likely a lensed event because it includes $p\!\geq\!0.05$ where the unlensed hypothesis is true. Third, for a cross-verification, we look at the Bayes factor $\mathcal{B}^\mathrm{ML}_\mathrm{U}$ from \cite{LIGOScientific:2021izm} and find that $\mathcal{B}^\mathrm{ML}_\mathrm{U}$ of GW190707\_093326 also disfavors lensing. Therefore, we conclude the signal of GW190707\_093326 is likely an unlensed signal and, consequently, we find no certain evidence of beating patterns from all evaluated BBH events as consistent as the observation made in the Bayes factor-based searches~\citep{Hannuksela:2019kle, LIGOScientific:2021izm}.

We organize this paper as follows: we describe the utilization of deep learning implemented in this search in Section~\ref{sec:method}, from the configuration of training data to the application of the ensemble learning-based majority voting strategy. In Section~\ref{sec:results}, we present the search results of the event classification. Then, we provide the summary and outlook of the search in Section~\ref{sec:conclusion}.

\begin{table*}[t!]
\caption{Parameters used for preparing simulated unlensed and lensed GW signals. The range of network $\mathrm{S/N}$ in the seventh row is the criterion for taking or discarding generated samples with the randomized parameters. $\mu_\pm$ and $\Delta t$ in the last three columns are the resulting values from the chosen $M_\mathrm{L}$, $D_\mathrm{L}$, and $D_\mathrm{LS}$. The prior distributions are set to be the same as those in \cite{Kim:2020xkm}.}
\centering
\begin{tabularx}{1\linewidth}{@{} *{2}{l} @{\extracolsep{\fill}} *{2}{c} @{}}
\toprule\toprule
Signal Type & Parameter & Range & Distribution\\
\hline
\multirow{6}{*}{$h_\mathrm{U}(f)$ \& $h_\mathrm{L}(f)$} & Component masses of source, $m_1$ \& $m_2$ & $5$ -- $100\msun$ & log-uniform\\
& Displacement of source, $\delta$ & $10^{-6}$ -- $0.5$ pc & uniform\\
& Right ascension of source location& $0^\circ$ -- $360^\circ$ & uniform\\
& Declination of source location& $\text{-}90^\circ$ -- $+90^\circ$ & uniform \\
& Polarization angle, $\psi$ & $0^\circ$ -- $360^\circ$ & uniform \\
& Inclination angle, $\iota$ & $0^\circ$ -- $360^\circ$ & uniform \\
\cline{2-4}
& Network signal-to-noise ratio, $\mathrm{S/N}_\textrm{net}$ & $10$ -- $50$ & \NA \\
\hline
\multirow{6}{*}{$h_\mathrm{L}(f)$} & Lens mass, $M_\mathrm{L}$ & $10^3$ -- $10^5\msun$ & log-uniform\\
& Distance to lens, $D_\mathrm{L}$ & $10$ -- $10^3$ Mpc & uniform \\
& Distance from lens to source, $D_\mathrm{LS}$ & $10$ -- $10^3$ Mpc & uniform\\
\cline{2-4}
& Magnification factors, $\mu_\pm$ & $\mu_+$: $1.17$ -- $10.51$; $\mu_-$: $\text{-}9.51$ -- $\text{-}0.17$ & \NA \\
& Time delay, $\Delta t$ & $2.25$ ms -- $3.52$ s & \NA \\
\bottomrule\bottomrule
\end{tabularx}
\label{tab:params_setup}
\end{table*}


\section{Implementation of Deep Learning}
\label{sec:method}

We implement one of the state-of-the-art deep learning models, VGG-19 (VGG hereafter) via \textsc{PyTorch}~\citep{NEURIPS2019_9015}, for the identification of the GW microlensing signature. In~\cite{Kim:2020xkm}, we have already shown the classification performance of the VGG model for distinguishing simulated lensed GW signals from unlensed ones is quite feasible: For example, the true positive rates of classifying lensed signals are $\gtrsim 97\%$ for all considered cases.

However, it is less proper to directly use the pre-trained model of the previous work because some of the considerations for preparing the training data were rather insufficient to reflect the reality. For example, we supposed (i) the face-on orientation for BBH merger systems and (ii) the optimal signal-to-noise ratio ($\mathrm{S/N}$) for GW signals with a simpler noise model, the Detuned High Power model of the Advanced LIGO~\citep{ZeroDetHighP} in the earlier work. Therefore, to mimic actual observations as much as possible, we partially change the parameter setups for the training data preparation and summarize the parameter setup used in this search in Table~\ref{tab:params_setup}.  Then we train the VGG model again with newly prepared training data.

On the other hand, it is known in general that a single learner of any machine learning model may result in an improperly biased prediction and, for such concerns, ensemble learning can be a prescription~\citep{10.5555/3153997}. To this end, we introduce an ensemble learning for this search. Then, we build a majority voting-based classification strategy for the classes predicted by the ensemble learners.

\subsection{Preparation of Training Data}
\label{sec:data_prep}

We comprise the training data with spectrogram samples of mock GW signals of unlensed and lensed BBH events. First, for the generation of simulated signals, we take a similar parameter setup to that of~\cite{Kim:2020xkm}: For the nonprecessing unlensed GW signal in the frequency-domain, $h_\mathrm{U}(f)$, we use the IMRPhenomPv2 model~\citep{Hannam:2013oca,Schmidt:2014iyl} with the parameters summarized in Table~\ref{tab:params_setup}. We adopt the same prior distributions of \cite{Kim:2020xkm}, i.e., the log-uniform population for the component masses and lens mass, and the uniform population for all other parameters without regarding specific prior models for the populations.\footnote{One can refer \cite{LIGOScientific:2021psn} for some practical prior models inferred from the observed GW events for interest.}

We suppose the thin-lens approximation for the lensed signals. Under the approximation, the lensed GW signal in the frequency-domain, $h_\mathrm{L}(f)$, can be described as follows:
\begin{equation}
h_\mathrm{L}(f) = F(f) h_\mathrm{U}(f)~, \label{eq:lensed_hf}
\end{equation}
where $F(f)$ is the amplification factor in the frequency-domain that determines GW lensing signatures.
For the typical mass for GW microlensing, $\lesssim\!10^5$\msun, taking into account $F(f)$ in the wave optics is a conventional treatment (e.g, \cite{Diego:2019lcd, Seo_microlensing}). However, \cite{takahashi:2003apj} had shown that $|F(f)|$ in the wave optics asymptotically converge to the geometrical optics limit when a dimensionless frequency $\omega \equiv 8\pi G M_{\mathrm{L}z} f / c^3 \gtrsim 1$. The condition is converted to $f \gtrsim\!0.1~\textrm{Hz}$ for the considered ranges of the redshifted lens mass $M_{\mathrm{L}z} = M_\mathrm{L}(1+z)$ in this search; we see the sensitive frequency band, $\sim\!10~\textrm{Hz}$--$1000~\textrm{Hz}$, of the ground-based detectors, corresponds to where the geometrical optic limit is valid. Hence, we take the analytic form of $F(f)$ in the geometrical optics limit provided in~\cite{takahashi:2003apj}:
\begin{equation}
F(f) = \sqrt{|\mu_{+}|} - i \sqrt{|\mu_{-}|} e^{2\pi i f \Delta t}~. \label{eq:F(f)_pm}
\end{equation}
Here, $\mu_\pm$ are the magnification factors of two lensed signals and $\Delta t$ is the time delay of arrival times between them. To compute $\mu_\pm$ and $\Delta t$, we additionally suppose point-like lenses: For the point-mass lens model, $\mu_\pm$ and $\Delta t$ are given as
\begin{eqnarray}
\mu_\pm &=& \frac{1}{2} \pm \frac{y^2 + 2}{2y\sqrt{y^2 + 4}}~, \label{eq:mu_pm}\\
\Delta t &=& \frac{4 G M_{\mathrm{L}z}}{c^3} \left[ \frac{y \sqrt{y^2 + 4}}{2} + \ln \left\{\frac{\sqrt{y^2+4}+y}{\sqrt{y^2+4}-y} \right\} \right]~, \nonumber \\
\label{eq:dt_pm}
\end{eqnarray}
respectively, where $y = (\delta D_{\mathrm{L}}) / (\xi_0 D_{\mathrm{S}})$ is a position parameter for source which is determined by the displacement of source, $\delta$, the Einstein radius, $\xi_0 = \sqrt{(4 G M_\mathrm{L} / c^2) D_{\mathrm{LS}} D_\mathrm{L} / D_\mathrm{S}}$ of a point-mass lens, along with the angular distances from observer to lens, $D_\mathrm{L}$, to source, $D_\mathrm{S}$, and between the lens and source, $D_{\mathrm{LS}}$. Now we can determine $F(f)$ by computing $\mu_\pm$ and $\Delta t$ with the parameters in Table~\ref{tab:params_setup}; the computed values of $\mu_+$, $\mu_-$, and $\Delta t$ are distributed within [$1.17, 10.51$], [$-9.51, -0.17$], and [$2.25~\mathrm{ms}, 3.52~\mathrm{s}$], respectively.

For the three observing runs, sensitivities of the Advanced LIGO and Virgo detectors were gradually enhanced (e.g., Figure 1 of \cite{KAGRA:2013rdx}) and it made the forty-six BBH events being observed in slightly different environments. Thus, we adopt the power spectral density of the Advanced LIGO's design sensitivity~\citep{Harry:2010zz} not only to regard a commonly applicable noise model for the target BBH events observed in non-identical environments of detectors and observing runs but also to mimic the non-removable noise represented in the spectrogram.

We inject the simulated signals into the noise curve data acquired from the \textsc{pycbc.psd} module of the \textsc{PyCBC} package~\citep{Usman:2015kfa, alex_nitz_2020_4075326}. We also constraint $\mathrm{S/N}$ of each signal similar to \cite{Kim:2020xkm}: In this search, we consider the network $\mathrm{S/N}$, $\mathrm{S/N}_\textrm{net}$, defined as
\begin{equation}
\mathrm{S/N}_\textrm{net} \equiv \left\{ \sum_\textrm{IFO} \mathrm{S/N}_\textrm{IFO}^2 \right\}^{1/2}~,
\end{equation}
where IFO denotes the LIGO-Hanford (H1), LIGO-Livingston (L1), and Virgo (V1) detectors and let the training data consist of spectrogram samples satisfying $10 \leq \mathrm{S/N}_\textrm{net} \leq 50$.\footnote{The reader can refer Section 3.1.3 of~\cite{Kim:2020xkm} for the discussion about the range.} 
By constraining $\mathrm{S/N}_\textrm{net}$, our training samples can be prepared fairly for the evaluation on the BBH events which were identified by the detection criterion $\mathrm{S/N}_\textrm{net} \geq 10$ at different sensitivities of the LIGO and Virgo detectors operated over the three observing runs.

We apply the constant-$Q$ transformation technique \citep{Chatterji:2004qg} via \textsc{pycbc.filter.qtransform} function to the noise-added $h_\mathrm{U}(f)$ and $h_\mathrm{L}(f)$ signals to generate the spectrogram samples.\footnote{For the transformation, we set the frequency range as [$20~\mathrm{Hz}, 350~\mathrm{Hz}$], $Q$-value range as [$16,16$], the step-size of time as 1/256, and the step-size of frequency as $\log_{10}256$. Note that the values are optimal choices determined empirically in this search.} However, we see from the observations on BBH events that the duration time of BBH signals spans less than $1$~second (for example, see Figure~10 of~\cite{LIGOScientific:2018mvr}) within the sensitive frequency band of the Advanced LIGO and Advanced Virgo detectors. Therefore, we trim the time window of the spectrograms to $[-0.9 ~\mathrm{s},+0.1~\mathrm{s}]$ around the event time of each event to enhance the search accuracy. By doing so, we can save in computational expenses additionally. 

We prepare 45,000 spectrogram samples for each of lensed and unlensed classes; We configure three independent subsets--training, development, and testing data---with randomly chosen 80\%, 10\%, and 10\% of total samples, respectively, and make each subset to contain the same number of lensed and unlensed samples. We normalize the pixel values of the spectrograms to $[-1,1]$ with the Min-Max normalization over all samples of both classes before feeding the data into VGG.

\subsection{Ensemble Learning}

For the implementation of ensemble learning, we repeat the training ten times with randomly chosen ten different random seeds for each training. Instead, we train all learners with the same training data by adopting the same training scheme built in the previous work~\citep{Kim:2020xkm}: The batch size is set to be 128 and the maximum training epoch is set as 100. We use the Adam optimization algorithm \citep{2014ADAM} to optimize the training accuracy. We make the trained model returns a probability, $r$, to the lensed class.

The loss functions taken for the error measurement $E$ of the training is the cross-entropy function
\begin{equation}
E = -r_i \log{\hat{r}_i}~, \label{eqn:cross_entropy}
\end{equation}
where $r_i$ and $\hat{r}_i$ are the target probability and the predicted probability of an $i$-th training sample, respectively. The training of the ten independent learners is conducted on an NVIDIA Tesla P40 GPU. For more details, one can refer~\cite{Kim:2020xkm}. We present the result of performance test on the ensemble learning in Appendix~\ref{apx:performance_test}.

\subsection{Majority Voting and Consistency-based Classification}
\label{sec:clf}

\begin{figure}[t!]
\centering
    \includegraphics[width=1.\linewidth]{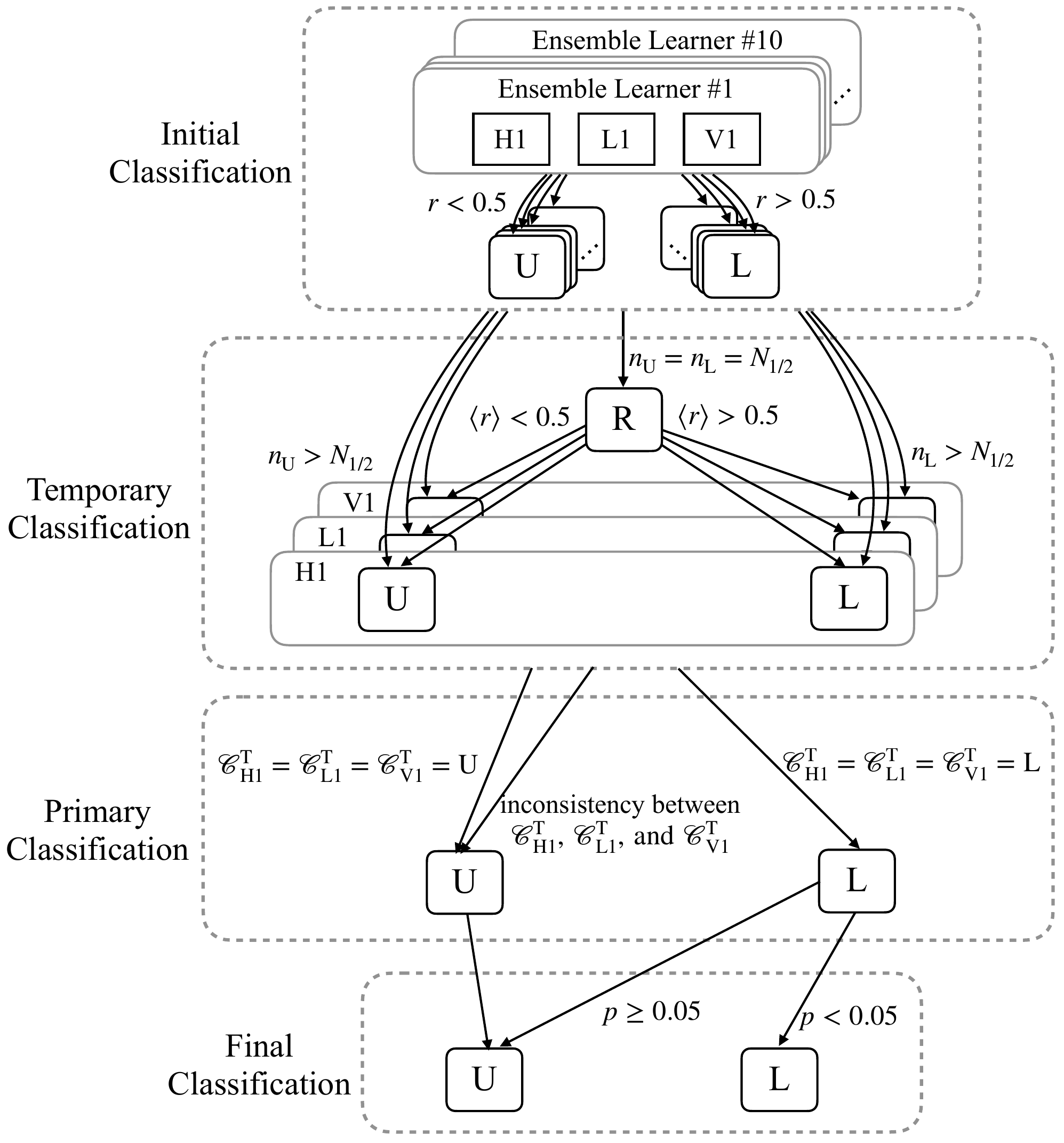}
\caption{Hierarchical classification scheme implemented in this search. The probability $r$ obtained from the ensemble learners is used in the initial classification. For temporary classification, we implement majority voting on the result of initial classification. If the majority voting meets the criterion C2-3, we compute the mean probability $\left<r\right>$ and judge the temporary class of each detector's data. We apply the consistency test on the result of temporary classification for the primary classification. Lastly, for the estimation of $p$-value in the final classification, we use the median probability and its 90\% credible interval obtained by bootstrapping the probabilities used in the initial classification. \label{fig:clf_scheme}}
\end{figure}

To determine the initial, temporary, primary, and final classes of an event, we apply a majority voting-aided consistency test which is deployed by following hierarchical manner (see also Figure~\ref{fig:clf_scheme}) with the given criteria (C):
\begin{quote}
\begin{itemize}[leftmargin=3mm, rightmargin=-7mm]
    \item[Step 1.] Determine an initial class ($\mathcal{C}^\mathrm{I}$) for the output of each ensember learner for each detector's data based on the probability\footnote{The probability is obtainable from the softmax activation function of the VGG model.} $r$ to the lensed class such that
    \begin{itemize}[leftmargin=2mm]
        \item[C1-1.] $\mathcal{C}^\mathrm{I} = \mathrm{U}$ if $r < 0.5$.
        \item[C1-2.] $\mathcal{C}^\mathrm{I} = \mathrm{L}$ if $r > 0.5$.
    \end{itemize}
    \item[Step 2.] Decide the temporary class ($\mathcal{C}^\mathrm{T}$) of each detector's data based on the majority voting based on the number of initial U classes ($n_\mathrm{U}$) and the number of initial L classes ($n_\mathrm{L}$) compared to the half of the total number of initial classes ($N_{1/2}$) such that
    \begin{itemize}[leftmargin=2mm]
        \item[C2-1.] $\mathcal{C}^\mathrm{T} = \mathrm{U}$ if $n_\mathrm{U} > N_{1/2}$.
        \item[C2-2.] $\mathcal{C}^\mathrm{T} = \mathrm{L}$ if $n_\mathrm{L} > N_{1/2}$.
        \item[C2-3.] $\mathcal{C}^\mathrm{T} = \mathrm{R}$ if $n_\mathrm{U} = n_\mathrm{L} = N_{1/2}$.
        \begin{itemize}[leftmargin=8mm]
            \item[C2-3-1.] $\mathcal{C}^\mathrm{T} = \mathrm{U}$ if $\left<r\right> < 0.5$.
            \item[C2-3-2.] $\mathcal{C}^\mathrm{T} = \mathrm{L}$ if $\left<r\right> > 0.5$.
        \end{itemize}
    \end{itemize}
    \item[Step 3.] Judge the primary class ($\mathcal{C}^\mathrm{P}$) of an event based on the consistency between $\mathcal{C}^\mathrm{T}$ of each detector's data such that
    \begin{itemize}[leftmargin=2mm]
        \item[C3-1.] $\mathcal{C}^\mathrm{P} = \mathrm{U}$ if $\mathcal{C}^\mathrm{T}_\mathrm{H1} = \mathcal{C}^\mathrm{T}_\mathrm{L1} = \mathcal{C}^\mathrm{T}_\mathrm{V1} = \mathrm{U}$.
        \item[C3-2.] $\mathcal{C}^\mathrm{P} = \mathrm{U}$ if $\mathcal{C}^\mathrm{T}_\mathrm{H1}$, $\mathcal{C}^\mathrm{T}_\mathrm{L1}$, and $\mathcal{C}^\mathrm{T}_\mathrm{V1}$ are inconsistent.
        \item[C3-3.] $\mathcal{C}^\mathrm{P} = \mathrm{L}$ if $\mathcal{C}^\mathrm{T}_\mathrm{H1} = \mathcal{C}^\mathrm{T}_\mathrm{L1} = \mathcal{C}^\mathrm{T}_\mathrm{V1} = \mathrm{L}$.
    \end{itemize}
    \item[Step 4.] Conclude the final class ($\mathcal{C}^\mathrm{F}$) of an event based on $C^\mathrm{P}$ and $p$-value such that
    \begin{itemize}[leftmargin=2mm]
        \item[C4-1.] $\mathcal{C}^\mathrm{F} = \mathrm{U}$ if $\mathcal{C}^\mathrm{P} = \mathrm{U}$.
        \item[C4-2.] $\mathcal{C}^\mathrm{F} = \mathrm{U}$ if $\mathcal{C}^\mathrm{P} = \mathrm{L}$ but $p \geq 0.05$.
        \item[C4-3.] $\mathcal{C}^\mathrm{F} = \mathrm{L}$ if $\mathcal{C}^\mathrm{P} = \mathrm{L}$ but $p < 0.05$.
    \end{itemize}
\end{itemize}
\end{quote}
Note that, for C3-2, we conservatively judge an event as U if $\mathcal{C}^\mathrm{T}$s are inconsistent with each other. Furthermore, for C4-1, we conclude $\mathcal{C}^\mathrm{F}$ of an event as U if $\mathcal{C}^\mathrm{P}=\mathrm{U}$ because it is not a new discovery. On the other hand, for a detector's data classified as R, i.e., reserved for a follow-up analysis from C2-3, we examine the mean probability $\left<r\right>$ to determine $\mathcal{C}^\mathrm{T}$ as either U or L via C2-3-1 or C2-3-2, respectively.

In particular, if the $\mathcal{C}^\mathrm{P}$ of an event is judged as L from C3-3, we verify the potential new discovery by estimating the $p$-value based on the model established in Section~\ref{sec:pval_model} for the probability on the event. Finally, we conclude the final class of the event via C4-2 or C4-3 according to the estimated $p$-value.

\subsection{Model for $p$-value Estimation}
\label{sec:pval_model}

We build a $p$-value model from the performance test on the testing data in order to estimate the confidence of the primary classification. For the computation of $p$-value, we use following equation:
\begin{equation}
p = 1 - \exp^{-N\mathcal{F}}~,
\end{equation}
where $N$ denotes the number of candidates satisfying the condition given in computing the false alarm probability $\mathcal{F}$ defined as 
\begin{equation}
\mathcal{F} = P(r^* \geq r^\mathrm{t} | \mathrm{U})~. \label{eq:fap}
\end{equation}
$\mathcal{F}$ means the probability of finding one or more samples having $r^*$s greater than or equal to $r^\mathrm{t}$---the probability of a target sample in the testing data---from the opposite class, i.e., the unlensed class $\mathrm{U}$.

\begin{figure}[t!]
\centering
\includegraphics[width=1\linewidth]{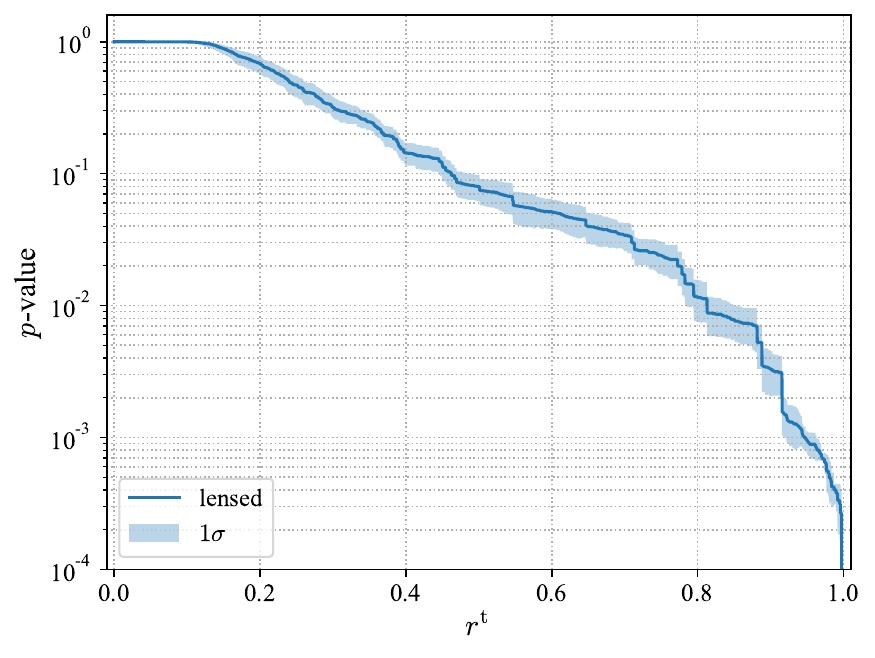}
\caption{$p$-value model built with respect to threshold probability for the lensed samples. The shaded region around the curves shows the $1\sigma$ uncertainty on the estimated $p$-value. \label{fig:pvalue}}
\end{figure}

We present the $p$-value model built with respect to $r^\mathrm{t}$ in Figure~\ref{fig:pvalue}. Note that we provide $1\sigma$ uncertainty of $p$-value that come from the different estimations of the ensemble learners with shaded regions. From this figure, we observe that the $p$-value becomes $<\!0.05$ when $r^\mathrm{t} \gtrsim 0.6$. Therefore, we set $r > 0.6$ and $p < 0.05$ as the empirical criterion for claiming the detection of a lensed signal.

\section{Search Results}
\label{sec:results}

\begin{table*}[t!]
\caption{Result of the initial classification for the each detector's data of the 46 BBH events in GWTC-1 and -2. The labels, U and \bL, indicate that the event is classified as \emph{unlensed} signal and \emph{lensed} signal, respectively. Note that ``Run \#'' in the header row denotes the index of each learner. We annotate the label of L class as \bL~to let them be easily distinguishable from the U class. We use this result to obtain the temporary class summarized in Table~\ref{tab:clf_result} or Table~\ref{tab:temp_class}. \label{tab:clf_result_ifo}}
\vspace{-4mm}
\begin{center}
\begin{tabularx}{1\linewidth}{@{} p{0.17\textwidth} *{3}{>{\centering\arraybackslash}p{0.008\textwidth}}
*{3}{>{\centering\arraybackslash}p{0.008\textwidth}}
*{3}{>{\centering\arraybackslash}p{0.008\textwidth}}
*{3}{>{\centering\arraybackslash}p{0.008\textwidth}}
*{3}{>{\centering\arraybackslash}p{0.008\textwidth}}
*{3}{>{\centering\arraybackslash}p{0.008\textwidth}}
*{3}{>{\centering\arraybackslash}p{0.008\textwidth}}
*{3}{>{\centering\arraybackslash}p{0.008\textwidth}}
*{3}{>{\centering\arraybackslash}p{0.008\textwidth}}
*{3}{>{\centering\arraybackslash}p{0.008\textwidth}}
@{}}
\toprule\toprule
& \multicolumn{3}{c}{Run~\#1} & \multicolumn{3}{c}{Run~\#2} & \multicolumn{3}{c}{Run~\#3} & \multicolumn{3}{c}{Run~\#4} & \multicolumn{3}{c}{Run~\#5} & \multicolumn{3}{c}{Run~\#6} & \multicolumn{3}{c}{Run~\#7} & \multicolumn{3}{c}{Run~\#8} & \multicolumn{3}{c}{Run~\#9} & \multicolumn{3}{c}{Run~\#10}\\[-1.6mm]
Events & H1 & L1 & V1 & H1 & L1 & V1 & H1 & L1 & V1 & H1 & L1 & V1 & H1 & L1 & V1 & H1 & L1 & V1 & H1 & L1 & V1 & H1 & L1 & V1 & H1 & L1 & V1 & H1 & L1 & V1\\
\hline
GW150914 & U & U & \NA & U & U & \NA & U & U & \NA & U & U & \NA & U & U & \NA & U & U & \NA & U & U & \NA & U & U & \NA & U & U & \NA & U & U & \NA\\[\colsep]
GW151012 & U & \bL & \NA & U & \bL & \NA & U & \bL & \NA & U & \bL & \NA & U & \bL & \NA & U & \bL & \NA & U & \bL & \NA & U & \bL & \NA & U & \bL & \NA & U & \bL & \NA\\[\colsep]
GW151226 & U & U & \NA & U & U & \NA & U & U & \NA & U & U & \NA & U & U & \NA & U & U & \NA & U & U & \NA & U & U & \NA & U & U & \NA & U & U & \NA\\[\colsep]
GW170104 & U & U & \NA & U & U & \NA & U & U & \NA & U & U & \NA & U & U & \NA & U & U & \NA & U & U & \NA & U & U & \NA & U & U & \NA & U & U & \NA\\[\colsep]
GW170608 & U & \bL & \NA & U & U & \NA & U & U & \NA & U & \bL & \NA & U & U & \NA & U & \bL & \NA & U & U & \NA & U & \bL & \NA & U & \bL & \NA & U & U & \NA\\[\colsep]
GW170729 & U & U & U & U & U & U & U & U & U & U & U & U & U & U & U & U & U & U & U & U & U & U & U & U & U & U & U & U & U & U\\[\colsep]
GW170809 & U & U & U & U & U & U & U & U & U & U & U & U & U & U & U & U & U & U & U & U & U & U & U & U & U & U & U & U & U & U\\[\colsep]
GW170814 & U & U & \bL & U & U & U & U & U & U & U & U & U & U & U & \bL & U & U & U & U & U & U & U & U & U & U & U & U & U & U & U\\[\colsep]
GW170818 & U & U & U & U & U & U & U & U & U & U & U & U & U & U & U & U & U & U & U & U & U & U & U & U & U & U & U & U & U & U\\[\colsep]
GW170823 & U & U & \NA & U & U & \NA & U & U & \NA & U & U & \NA & U & U & \NA & U & U & \NA & U & U & \NA & U & U & \NA & U & U & \NA & U & U & \NA\\[\colsep]
GW190408\_181802 & U & U & U & U & U & U & U & U & U & U & U & U & U & U & U & U & U & U & U & U & U & U & \bL & U & U & U & U & U & U & U\\[\colsep]
GW190412 & U & \bL & U & U & \bL & U & U & \bL & U & U & \bL & U & U & U & U & U & \bL & U & U & U & U & U & \bL & U & U & \bL & U & U & \bL & U\\[\colsep]
GW190413\_052954 & U & \bL & U & U & \bL & U & U & \bL & U & U & \bL & U & U & \bL & \bL & U & \bL & U & U & \bL & U & U & U & U & U & \bL & U & U & \bL & U\\[\colsep]
GW190413\_134308 & U & U & U & U & U & U & U & U & U & U & U & U & U & U & U & U & U & U & U & U & \bL & U & U & U & U & U & U & U & U & U\\[\colsep]
GW190421\_213856 & U & U & \NA & U & U & \NA & U & U & \NA & U & U & \NA & U & U & \NA & U & U & \NA & U & U & \NA & U & U & \NA & U & U & \NA & U & U & \NA\\[\colsep]
GW190424\_180648 & \NA & U & \NA & \NA & U & \NA & \NA & U & \NA & \NA & U & \NA & \NA & U & \NA & \NA & U & \NA & \NA & U & \NA & \NA & U & \NA & \NA & U & \NA & \NA & U & \NA\\[\colsep]
GW190503\_185404 & U & U & U & U & U & U & U & U & U & U & U & U & U & U & U & U & U & U & U & U & U & U & U & U & U & U & U & U & U & U\\[\colsep]
GW190512\_180714 & U & U & \bL & U & U & \bL & U & U & U & U & U & U & U & U & U & U & U & U & U & U & \bL & U & U & U & U & U & \bL & U & U & U\\[\colsep]
GW190513\_205428 & U & \bL & U & U & \bL & U & U & \bL & U & U & \bL & U & U & \bL & U & U & \bL & U & U & \bL & U & U & \bL & U & U & \bL & U & U & \bL & U\\[\colsep]
GW190514\_065416 & U & U & \NA & U & U & \NA & U & U & \NA & U & U & \NA & U & U & \NA & U & U & \NA & U & U & \NA & U & U & \NA & U & U & \NA & U & U & \NA\\[\colsep]
GW190517\_055101 & U & U & U & U & U & \bL & U & U & \bL & U & U & U & U & U & U & U & U & \bL & U & U & U & U & U & U & U & U & \bL & U & U & U\\[\colsep]
GW190519\_153544 & U & U & U & U & U & U & U & U & U & U & U & U & U & U & U & U & U & U & U & U & U & U & U & U & U & U & U & U & U & U\\[\colsep]
GW190521 & U & U & U & U & U & U & U & U & U & U & U & U & U & U & U & U & U & U & U & U & U & U & U & \bL & U & U & U & U & U & U\\[\colsep]
GW190521\_074359 & U & U & \NA & U & U & \NA & U & U & \NA & U & U & \NA & U & U & \NA & U & U & \NA & U & U & \NA & U & U & \NA & U & U & \NA & U & U & \NA\\[\colsep]
GW190527\_092055 & \bL & U & \NA & \bL & U & \NA & \bL & U & \NA & \bL & U & \NA & \bL & U & \NA & \bL & U & \NA & \bL & U & \NA & U & \bL & \NA & \bL & U & \NA & \bL & U & \NA\\[\colsep]
GW190602\_175927 & U & U & \bL & U & U & U & U & U & U & U & U & U & U & U & U & U & U & U & U & U & U & U & U & U & U & U & U & U & U & U\\[\colsep]
GW190620\_030421 & \NA & U & U & \NA & U & \bL & \NA & U & U & \NA & U & U & \NA & U & \bL & \NA & U & \bL & \NA & U & \bL & \NA & U & \bL & \NA & U & \bL & \NA & U & U\\[\colsep]
GW190630\_185205 & \NA & U & U & \NA & U & U & \NA & U & U & \NA & U & U & \NA & U & U & \NA & U & U & \NA & U & U & \NA & U & U & \NA & U & U & \NA & U & U\\[\colsep]
GW190701\_203306 & U & U & U & U & U & U & U & U & U & U & U & U & U & U & U & U & U & U & U & U & U & U & U & U & U & U & U & U & U & U\\[\colsep]
GW190706\_222641 & U & U & U & U & U & U & U & U & U & U & U & U & U & U & U & U & U & U & U & U & \bL & U & U & \bL & U & U & U & U & U & U\\[\colsep]
GW190707\_093326 & \bL & \bL & \NA & \bL & \bL & \NA & \bL & \bL & \NA & U & \bL & \NA & \bL & \bL & \NA & \bL & \bL & \NA & U & \bL & \NA & U & \bL & \NA & U & \bL & \NA & U & \bL & \NA\\[\colsep]
GW190708\_232457 & \NA & U & \bL & \NA & U & \bL & \NA & U & \bL & \NA & U & \bL & \NA & U & U & \NA & U & U & \NA & U & \bL & \NA & U & U & \NA & U & U & \NA & U & U\\[\colsep]
GW190719\_215514 & U & U & \NA & U & U & \NA & U & U & \NA & U & U & \NA & U & U & \NA & U & U & \NA & U & U & \NA & U & U & \NA & U & U & \NA & U & U & \NA\\[\colsep]
GW190720\_000836 & U & \bL & U & U & \bL & U & U & \bL & U & U & U & U & U & \bL & U & U & U & U & U & U & U & U & U & U & U & U & U & U & U & U\\[\colsep]
GW190727\_060333 & U & U & U & U & U & U & U & U & U & U & U & U & U & U & U & U & U & U & U & U & U & U & U & U & U & U & U & U & U & U\\[\colsep]
GW190728\_064510 & U & \bL & U & U & U & U & U & U & U & U & U & U & U & U & \bL & U & U & \bL & U & U & \bL & U & U & U & U & U & U & U & U & U\\[\colsep]
GW190731\_140936 & U & U & \NA & U & U & \NA & U & U & \NA & U & U & \NA & U & U & \NA & U & U & \NA & U & U & \NA & U & U & \NA & U & U & \NA & U & U & \NA\\[\colsep]
GW190803\_022701 & \bL & U & U & \bL & U & U & \bL & U & U & U & U & U & U & U & U & \bL & U & U & U & U & U & U & U & U & U & U & U & U & U & U\\[\colsep]
GW190828\_063405 & U & U & \bL & U & U & U & U & U & U & U & U & U & U & U & \bL & U & U & U & U & U & \bL & U & U & \bL & U & U & \bL & U & U & U\\[\colsep]
GW190828\_065509 & U & U & \bL & U & U & \bL & U & U & U & U & U & \bL & U & U & \bL & U & U & \bL & U & U & \bL & U & U & \bL & U & U & U & U & U & \bL\\[\colsep]
GW190909\_114149 & U & U & \NA & U & U & \NA & U & U & \NA & U & U & \NA & U & U & \NA & U & U & \NA & U & U & \NA & U & U & \NA & U & U & \NA & U & U & \NA\\[\colsep]
GW190910\_112807 & \NA & U & U & \NA & U & U & \NA & U & U & \NA & U & U & \NA & U & U & \NA & U & U & \NA & U & U & \NA & U & U & \NA & U & U & \NA & U & U\\[\colsep]
GW190915\_235702 & U & U & U & U & U & U & U & U & U & U & U & U & U & U & U & U & U & U & U & U & U & U & U & U & U & U & U & U & U & U\\[\colsep]
GW190924\_021846 & U & \bL & U & U & \bL & U & U & \bL & U & U & \bL & U & U & \bL & U & U & \bL & U & \bL & \bL & U & U & \bL & U & U & \bL & U & U & \bL & U\\[\colsep]
GW190929\_012149 & U & U & U & U & U & U & U & U & U & U & U & U & U & U & U & U & U & U & U & U & \bL & U & U & U & U & U & U & U & U & U\\[\colsep]
GW190930\_133541 & \bL & U & \NA & \bL & U & \NA & \bL & U & \NA & U & U & \NA & \bL & U & \NA & U & U & \NA & \bL & U & \NA & U & U & \NA & U & U & \NA & U & U & \NA\\
\bottomrule\bottomrule
\end{tabularx}
\end{center}
\end{table*}

\begin{table*}[t!]
\caption{Result of the event classification for the 46 BBH events in GWTC-1 and -2. The labels, U and L, indicate that the event is classified as \emph{unlensed} signal and \emph{lensed} signal, respectively. Note that the class shown as L$^*$ denotes the confirmed temporary class which was originally marked as R according to the criterion C2-3 of the classification scheme. From the list of primary class, we see that GW190707\_093326 is classified as lensed while all others are classified as unlensed from either C3-1 or C3-2. However, from the follow-up analyses, we conclude that the event is likely an unlensed one. Thus, the final class of GW190707\_093326 is marked as U. \label{tab:clf_result}}
\vspace{-3mm}
\begin{center}
\begin{tabularx}{1.\linewidth}{@{} p{0.13\textwidth} C{1mm} C{1mm} C{1mm} C{7.5mm} C{6mm} | p{0.13\textwidth} C{1mm} C{1mm} C{1mm} C{7.5mm} C{6mm} | p{0.13\textwidth} C{1mm} C{1mm} C{1mm} C{7.5mm} C{6mm} @{}}
\toprule\toprule
\multirow{3}{*}{Event} & \multicolumn{3}{C{3mm}}{\centering Temporary} & \multirow{3}{7.5mm}{\centering Primary Class} & \multirow{3}{6mm}{\centering Final Class} & \multirow{3}{*}{Event} & \multicolumn{3}{C{3mm}}{\centering Temporary} & \multirow{3}{7.5mm}{\centering Primary Class} & \multirow{3}{6mm}{\centering Final Class} & \multirow{3}{*}{Event} & \multicolumn{3}{C{3mm}}{\centering Temporary} & \multirow{3}{7.5mm}{\centering Primary Class} & \multirow{3}{6mm}{\centering Final Class}\\[\colsep]
 & \multicolumn{3}{C{3mm}}{Class} & & & & \multicolumn{3}{C{3mm}}{Class} & & & & \multicolumn{3}{C{3mm}}{Class} & &\\
\cline{2-4}\cline{8-10}\cline{14-16}
 & H1 & L1 & V1 & & & & H1 & L1 & V1 & & & & H1 & L1 & V1 & & \\
\hline
GW150914 & U & U & \NA & U & U & GW190503\_185404 & U & U & U & U & U & GW190719\_215514 & U & U & \NA & U & U \\[\colsep]
GW151012 & U & L & \NA & U & U & GW190512\_180714 & U & U & U & U & U & GW190720\_000836 & U & U & U & U & U\\[\colsep]
GW151226 & U & U & \NA & U & U & GW190513\_205428 & U & L & U & U & U & GW190727\_060333 & U & U & U & U & U\\[\colsep]
GW170104 & U & U & \NA & U & U & GW190514\_065416 & U & U & \NA & U & U & GW190728\_064510 & U & U & U & U & U\\[\colsep]
GW170608 & U & L$^*$ & \NA & U & U & GW190517\_055101 & U & U & U & U & U & GW190731\_140936 & U & U & \NA & U & U\\[\colsep]
GW170729 & U & U & U & U & U & GW190519\_153544 & U & U & U & U & U & GW190803\_022701 & U & U & U & U & U\\[\colsep]
GW170809 & U & U & U & U & U & GW190521 & U & U & U & U & U & GW190828\_063405 & U & U & L$^*$ & U & U\\[\colsep]
GW170814 & U & U & U & U & U & GW190521\_074359 & U & U & \NA & U & U &  GW190828\_065509 & U & U & L & U & U\\[\colsep]
GW170818 & U & U & U & U & U & GW190527\_092055 & L & U & \NA & U & U & GW190909\_114149 & U & U & \NA & U & U\\[\colsep]
GW170823 & U & U & \NA & U & U & GW190602\_175927 & U & U & U & U & U & GW190910\_112807 & \NA & U & U & U & U\\[\colsep]
GW190408\_181802 & U & U & U & U & U & GW190620\_030421 & \NA & U & L & U & U & GW190915\_235702 & U & U & U & U & U\\[\colsep]
GW190412 & U & L & U & U & U & GW190630\_185205 & \NA & U & U & U & U & GW190924\_021846 & U & L & U & U & U\\[\colsep]
GW190413\_052954 & U & L & U & U & U & GW190701\_203306 & U & U & U & U & U & GW190929\_012149 & U & U & U & U & U\\[\colsep]
GW190413\_134308 & U & U & U & U & U & GW190706\_222641 & U & U & U & U & U & GW190930\_133541 & L$^*$ & U & \NA & U & U\\[\colsep]
GW190421\_213856 & U & U & \NA & U & U & GW190707\_093326 & L$^*$ & L & \NA & L & U & & & &\\[\colsep]
GW190424\_180648 & \NA & U & \NA & U & U & GW190708\_232457 & \NA & U & L$^*$ & U & U & & & &\\
\bottomrule\bottomrule
\end{tabularx}
\end{center}
\end{table*}

We present the results of the deep learning-based search for beating patterns from GW signals of the forty-six BBH events in GWTC-1 and -2. We take the public strain data (32 seconds-long and sampling rate of 4096 Hz) from the Gravitational Wave Open Science Center~\citep{LIGOScientific:2019lzm} to configure the evaluation data, i.e., spectrograms of the GW signals of those events. We implement the same constant-\emph{Q} transformation technique and the reduced time window rule applied to the configuration of training data to the strain data too. With this treatment, we can prepare fair spectrograms of the signals to the spectrograms of the training data.

We summarize the result of the initial classification in Table~\ref{tab:clf_result_ifo} obtained by the criteria C1-1 and C1-2. In Appendix~\ref{apx:result_initial}, we provide the initial probabilities predicted by the ensemble learners and, eventually, used for the initial classification. We tabulate the final class of each event in the last column of Table~\ref{tab:clf_result} together with the temporary classes obtained from the initial classes of the ensemble learners for each detector's data and the primary class obtained by the majority voting-based consistency test for each event. 

\begin{table}[t!]
\caption{Confirmation of temporary class for the reserved classifications. \label{tab:temp_class}}
\vspace{-4mm}
\begin{center}
\begin{tabularx}{1.\linewidth}{@{} l @{\extracolsep{\fill}} *{4}{c} @{}}
\toprule\toprule
\multirow{3}{*}{Event} & \multirow{3}{*}{IFO} & Original & \multirow{3}{*}{$\left< r \right>$} & Confirmed \\ [\colsep]
& & Temporary & & Temporary \\ [\colsep]
& & Class & & Class \\
\hline
GW170608 & L1 & R & 0.62 & L\\[\colsep]
GW190707\_093326 & H1 & R & 0.58 & L\\[\colsep]
GW190708\_232457 & V1 & R & 0.53 & L\\[\colsep]
GW190828\_063405 & V1 & R & 0.51 & L\\[\colsep]
GW190930\_133541 & H1 & R & 0.57 & L\\
\bottomrule\bottomrule
\end{tabularx}
\end{center}
\end{table}

From the initial classes tabulated in Table~\ref{tab:clf_result_ifo}, we find that 5 events, GW170608, GW190707\_093326, GW190708 \_232457, GW190828\_063405,~and~GW190930\_133541, contain data which meets the criterion C2-3, i.e., $n_\mathrm{U} = n_\mathrm{L} = N_{1/2}$, so  originally marked as R in the temporary classification. But we need the temporary class to be determined as either L or U, so we test their mean probabilities and confirm that their temporary classes are all L as shown in Table~\ref{tab:temp_class}. 

We see from the confirmed temporary classes and the primary classes that 14 events meet the criterion C2-2: among them, 13 events, GW151012, GW170608, GW190412, GW190413\_052954, GW190513\_205428, GW190527\_092055, GW190620\_030421, GW190708\_232457, GW190828\_063405, GW190828\_065509, GW190924\_021846, GW190929\_012149, and GW190930\_133541, are of C3-2 (inconsistent between the temporary classes) and the remaining 1 event, GW190707\_093326, is of C3-3 (all data are classified as lensed); other 32 events correspond to the criteria, C2-1 and C3-1, i.e., all available detectors' data are classified as U from the temporary classification and, eventually, the events are classified as U from the primary classifications. 

\begin{figure}[t!]
\centering
\includegraphics[width=1.\linewidth]{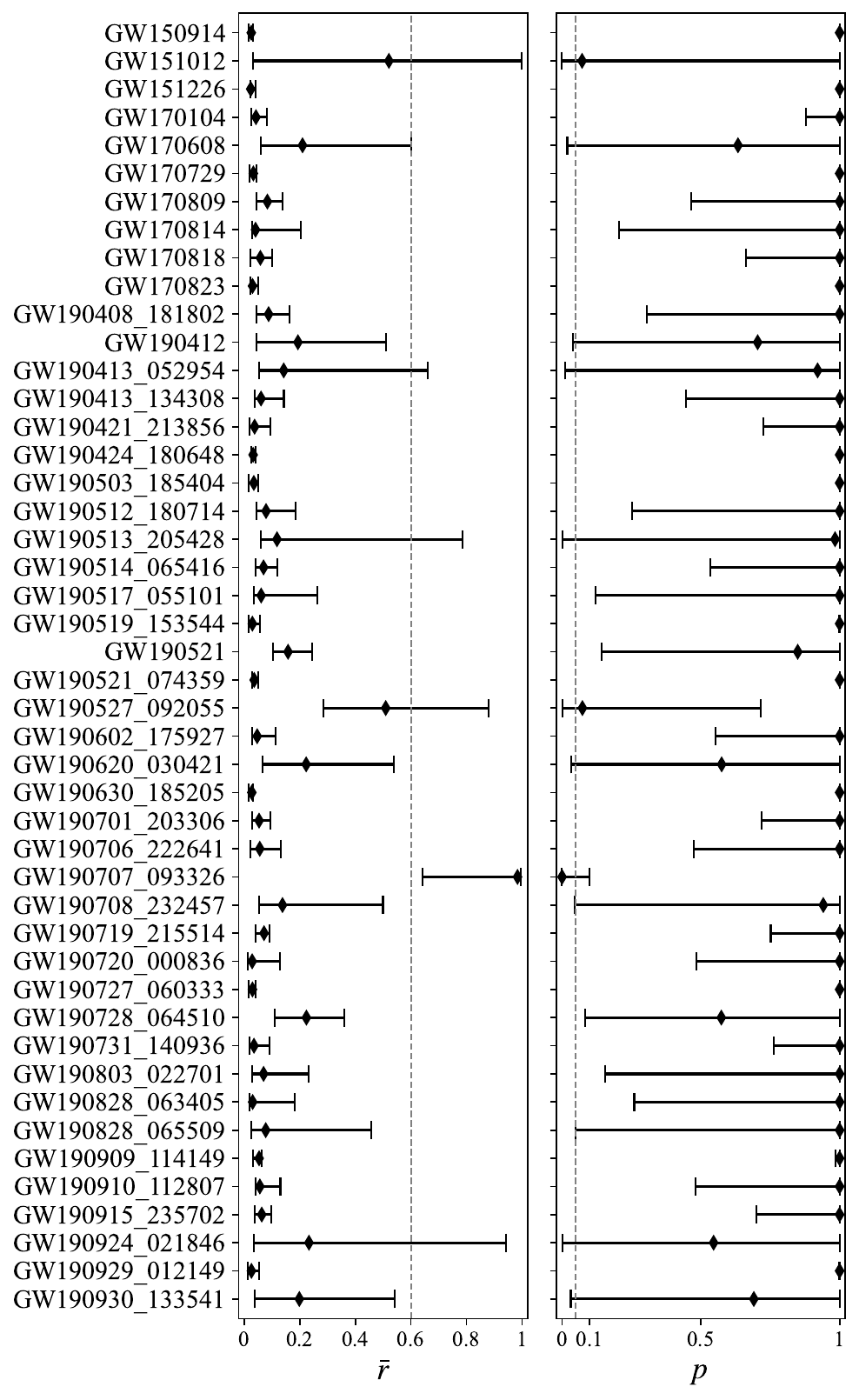}
\caption{Median probability $\bar{r}$ of the evaluated BBH events (left panel) and corresponding $p$-values (right panel). The marker in the left panel indicates the value of $\bar{r}$ and that in the right panel indicates the $p$-value corresponding to $\bar{r}$. The error bar on $\bar{r}$ shows 90\% CI while it on $p$ shows the lower and upper bounds of $1\sigma$ uncertainty for the 90\% CI of $\bar{r}$. The gray-dashed lines show the empirical threshold determined from the $p$-value model described in Section~\ref{sec:pval_model}. We can see that GW190707\_093326 is the only event satisfying $\bar{r}\!>\!0.6$ and $p(\bar{r})\!<\!0.05$. However, we observe that the uncertainty of its $p$-value is still including $p \geq 0.05$ which makes us to decide the event as likely an unlensed signal. \label{fig:probs_pvals}}
\end{figure}

For GW190707\_093326, the only event classified as L from the primary classification, we compute the median probability, $\bar{r}$, from the probabilities predicted by the ensemble learners. To compute $\bar{r}$, we introduce bootstrapping which iterates the ten probabilities 10,000 times and we obtain $\bar{r}=0.984^{+0.012}_{-0.342}$. Here, the error means the 90\% credible interval (CI) for $\bar{r}$.
We depict $\bar{r}$s of all other events with their 90\% CIs in the left panel of Figure~\ref{fig:probs_pvals}  together with that of GW190707\_093326 for comparison. We see that there are 6 events, GW151012, GW170608, GW190413\_052954, GW190513\_205428, GW190527\_092055, and GW190924\_021846 partially satisfying $\geq\!0.6$ although their $\bar{r}$s are computed below the empirical threshold. We observe that GW190707\_093326 is the only event of which the whole range of the uncertainty agrees to the criterion. We also find from this observation that the result of the majority voting-based consistency test is consistent with the probability-based classification.

We estimate the $p$-value of each event based on the model built in Section~\ref{sec:pval_model} regarding the uncertainty in $r$. We present the resulting $p$-values in the right panel of Figure~\ref{fig:probs_pvals}. From the estimated $p$-values and their uncertainties, we find that those 7 events satisfying $\bar{r}\!>0.6$ are including $p\!\geq\!0.05$, even for GW190707\_093326, convincing the unlensed hypothesis being true.

On the other hand, \cite{LIGOScientific:2021izm} have also searched beating patterns that might be occurred by microlenses with masses for $M_\mathrm{L} \lesssim 10^5$\msun~which are quite similar to the consideration of this work and the authors have defined the Bayes factor $\mathcal{B}^\mathrm{ML}_\mathrm{U}$ as an ultimate measure for testing the lensed hypothesis; it turned out that, $\log_{10} \mathcal{B}^\mathrm{ML}_\mathrm{U}$ of GW190707\_093326 is $\text{-}0.4$, i.e., the unlensed hypothesis is favored.

\begin{figure}[t!]
\centering
    \subfigure[GW190707\_093326 (LIGO-Livingston, L1)]
    {
        \includegraphics[width=1.\linewidth]{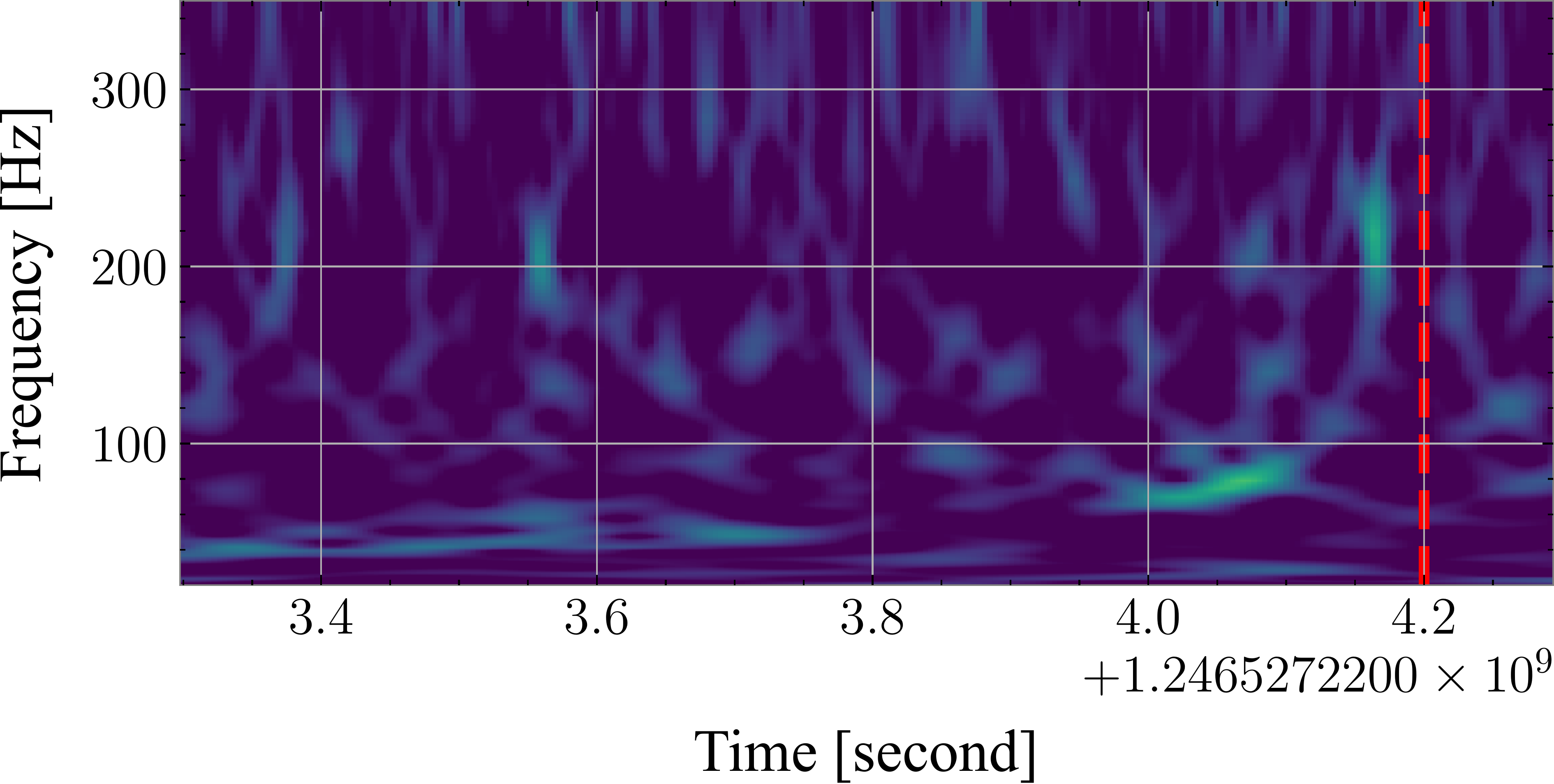}
    }\\
    \subfigure[GW190707\_093326 (LIGO-Hanford, H1)]
    {
        \includegraphics[width=1.\linewidth]{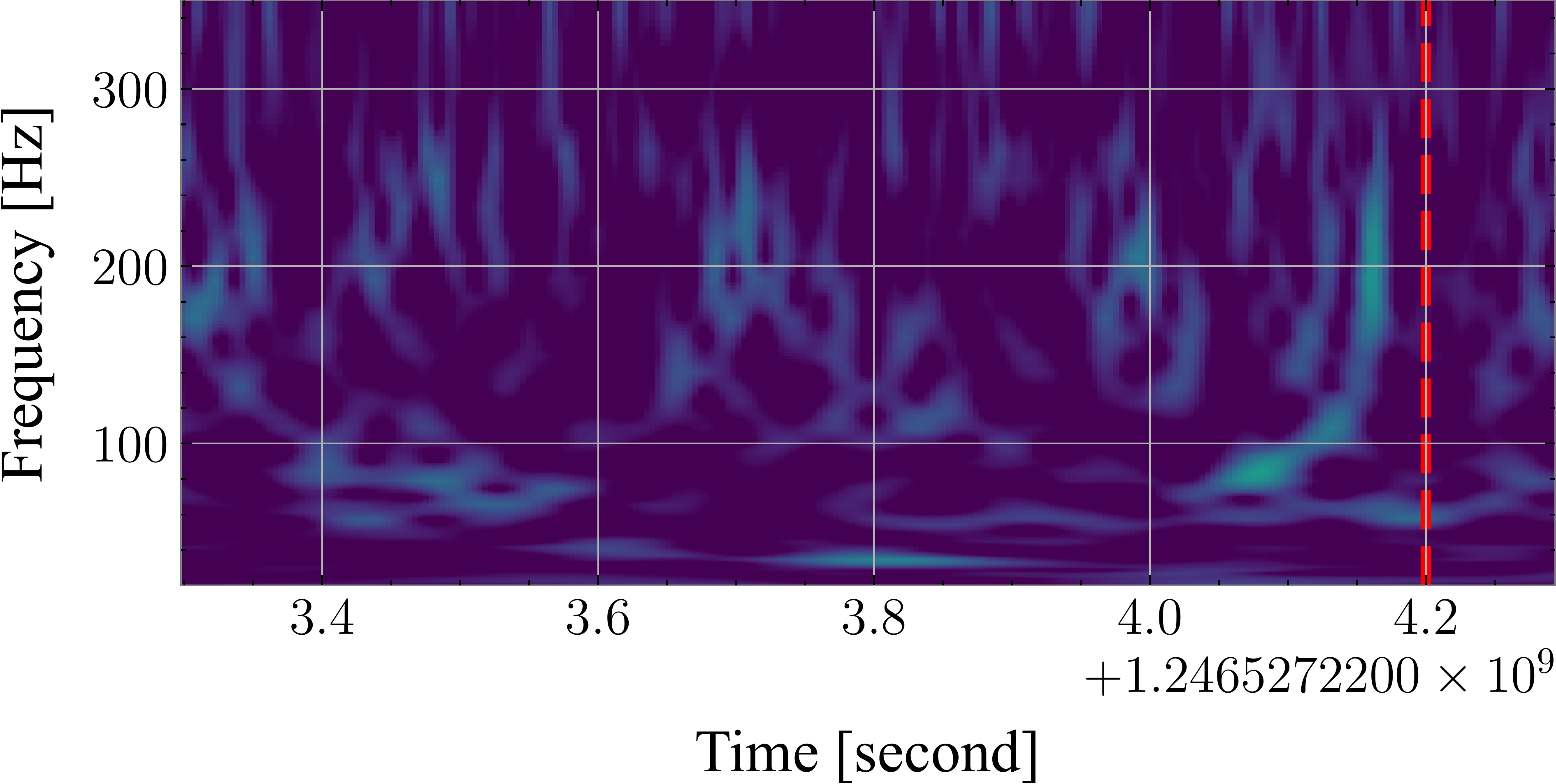}
    }\\
\caption{Evaluated spectrograms of GW190707\_093326 classified as lensed signal from the primary classification. The red-dashed line indicates the event time. Times in the horizontal axis are shown relative to the start time, $1246527220.0~\textrm{seconds}$ in the GPS time corresponding to July 7, 2019 at 09:33:22 UTC, of the data segment used in this search. We see no signature of beating patterns such as multiple sharp nodes or multiple peaks is shown from the chirp signal between [$4.0$, $4.2$] $\textrm{seconds}$ from the start time.\label{fig:misclf}}
\end{figure}

In addition to the quantitative analyses, we inspect the spectrogram of GW190707\_093326. In Figure~\ref{fig:misclf}, we present the evaluated spectrograms\footnote{The spectrograms are produced by the same manner for the preparation of training data described in Section~\ref{sec:data_prep} and they are slightly different from the ones which can be found from the Gravitational Wave Open Science Center.} of GW190707\_093326. Note that, for the event, only two LIGO detectors, LIGO-Hanford and LIGO-Livingston, provided available data. From the spectrogram, we see that no characteristic beating pattern, i.e., neither multiple peaks nor multiple sharp nodes that might come from the possible time delay by lenses with masses between $10^3$--$10^5$\msun, is shown. Particularly, even though we see from the spectrogram of LIGO-Livingston data of GW190707\_093326 (the top panel of Figure~\ref{fig:misclf}) that the energy, represented as the brightness, of the chirp signals changes as the time evolves compared to that of LIGO-Hanford data (the bottom panel of Figure~\ref{fig:misclf}), it is hard to confirm the beating pattern of lensed signals like the examples shown in~\cite{Kim:2020xkm}. 

We conclude from the follow-up analyses and a cross-verification with the Bayes factor that the signal of GW190707\_093326 is likely an unlensed signal as shown in the final class of the event in Table~\ref{tab:clf_result}. Therefore, in this search, we find no certain evidence of beating patterns from all evaluated BBH events as consistent as the conclusion of \cite{Hannuksela:2019kle} and \cite{LIGOScientific:2021izm}.


\section{Summary and Outlook}
\label{sec:conclusion}

We have presented the result of the first deep learning-based search for beating patterns from spectrograms of the gravitational-wave (GW) signals of the 46 binary black hole (BBH) events in the first and second GW transient catalogs. We have deployed a deep learning model, VGG-19, trained to seek beating patterns caused by lenses with masses between $10^3$--$10^5$\msun~from the BBH events in the catalogs. The majority voting and consistency-based primary classification predicted one event, GW190707\_093326, might be lensed.

However, we found that the uncertainty of the estimated $p$-value of GW190707\_093326 ranges from $0$ to $0.1$ for the 90\% confidence interval of the median probability to the lensed class and the range yet includes the possibility of the unlensed hypothesis being true, i.e., $p\!\geq\!0.05$. Therefore, we concluded that the signal of GW190707\_093326 is likely an unlensed one and found no significant evidence of beating patterns from the evaluated 46 BBH events.

For upcoming observing runs, we believe the method employed in this search will be a complementary tool to the existing methods, e.g., \cite{Lai:2018prd}, have been developed for searching beating patterns induced by the microlensing in GWs. To this end, it will be worth to consider some extended lens models such as singular isothermal sphere, signular isothermal ellipsoid, and so on. Moreover, adopting another waveform model such as IMRPhenomXPHM~\citep{Pratten:2020ceb} will make the method of this work to be consistent to the sophisticate state-of-the-art search pipelines summarized in~\cite{LIGOScientific:2021izm}.


\acknowledgements
KK thank Min-Su Shin for constructive discussion on the application of deep learning. We thank Srashti Goyal, Thomas Dent, Young-Min Kim, John J. Oh, SangHoon Oh, and Edwin J. Son for their fruitful comments on this work. We also thank the Global Science experimental Data hub Center (GSDC) at KISTI for supporting GPU-based computing resource. KK is supported by the National Research Foundation of Korea (NRF) grant funded by the Ministry of Science and ICT of the Korea Government (NRF-2020R1C1C1005863). OAH is supported by the research program of the Netherlands Organization for Scientific Research (NWO). TGFL is partially supported by grants from the Research Grants Council of Hong Kong (Project No. 14306218), Research Committee of the Chinese University of Hong Kong, and the Croucher Foundation of Hong Kong.

This research has made use of data or software obtained from the Gravitational Wave Open Science Center (gw-openscience.org), a service of LIGO Laboratory, the LIGO Scientific Collaboration, the Virgo Collaboration, and KAGRA. LIGO Laboratory and Advanced LIGO are funded by the United States National Science Foundation (NSF) as well as the Science and Technology Facilities Council (STFC) of the United Kingdom, the Max-Planck-Society (MPS), and the State of Niedersachsen/Germany for support of the construction of Advanced LIGO and construction and operation of the GEO600 detector. Additional support for Advanced LIGO was provided by the Australian Research Council. Virgo is funded, through the European Gravitational Observatory (EGO), by the French Centre National de Recherche Scientifique (CNRS), the Italian Istituto Nazionale di Fisica Nucleare (INFN) and the Dutch Nikhef, with contributions by institutions from Belgium, Germany, Greece, Hungary, Ireland, Japan, Monaco, Poland, Portugal, Spain. The construction and operation of KAGRA are funded by Ministry of Education, Culture, Sports, Science and Technology (MEXT), and Japan Society for the Promotion of Science (JSPS), National Research Foundation (NRF) and Ministry of Science and ICT (MSIT) in Korea, Academia Sinica (AS) and the Ministry of Science and Technology (MoST) in Taiwan.


\appendix

\section{Performance Test of Ensemble Learning}
\label{apx:performance_test}

\begin{figure}[t!]
\centering
\includegraphics[width=1\linewidth]{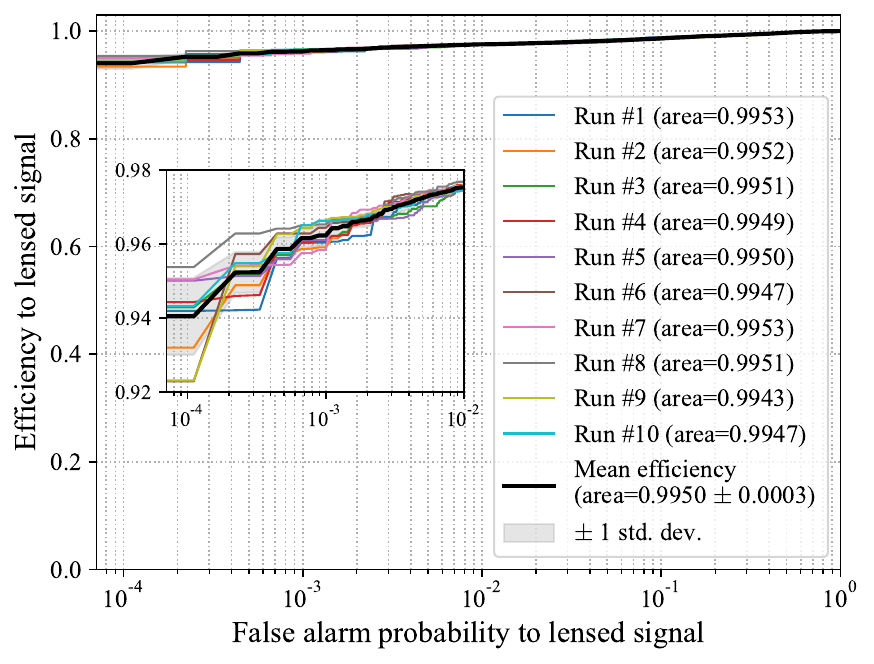}
\caption{Receiver operating characteristic (ROC) curves for lensed signal samples obtained by collecting the ROC curve of each learner. Again, ``Run \#'' indicates the index of each learner. We see only small deviations on the efficiency or the area under curve between the ensemble learners. But we observe that all ensemble learners can correctly distinguish lensed signals from unlensed ones with $>\!92\%$ efficiency even at the lowest false alarm probability. \label{fig:roc}}
\end{figure}

Figure~\ref{fig:roc} shows the receiver operating characteristic (ROC) curves obtained by evaluating testing samples of lensed class. Note that the horizontal and vertical axes are the false alarm probability $\mathcal{F}$ given in Equation~\eqref{eq:fap} and efficiency ($\mathcal{E}$) defined as
\begin{equation}
\mathcal{E} = P(r^* \geq r^\mathrm{t} | \mathrm{L})~, \label{eq:eff}
\end{equation}
respectively. Equation~\eqref{eq:eff} means the probability of finding one or more samples having $r^*$s greater than or equal to $r^\mathrm{t}$---the probability of a target lensed sample in the testing data---from samples of lensed class L. From the ROC curves, we see that all ensemble learners can mostly correctly classify testing samples from observing $\mathcal{E}\!>0.92$ and the area $>\!0.99$ for all curves.

\begin{figure*}[t!]
\centering
    \subfigure{
    \includegraphics[width=0.32\linewidth]{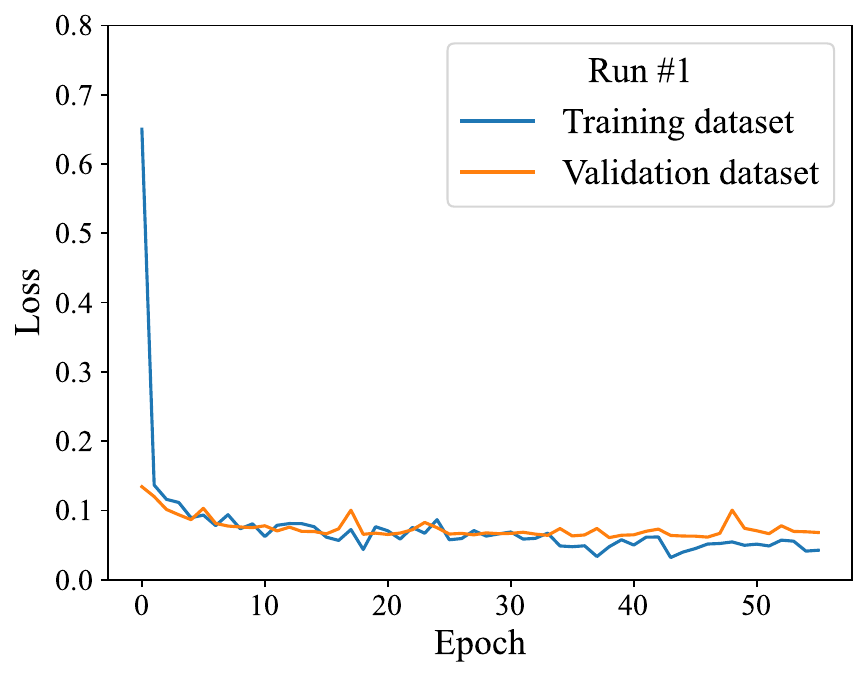}
    }
    \subfigure{
    \includegraphics[width=0.32\linewidth]{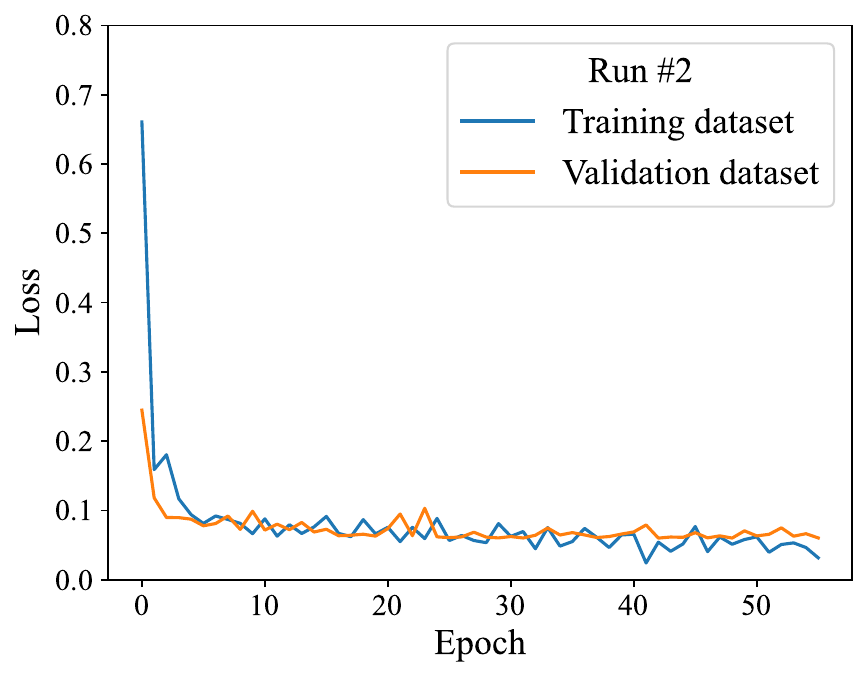}
    }
    \subfigure{
    \includegraphics[width=0.32\linewidth]{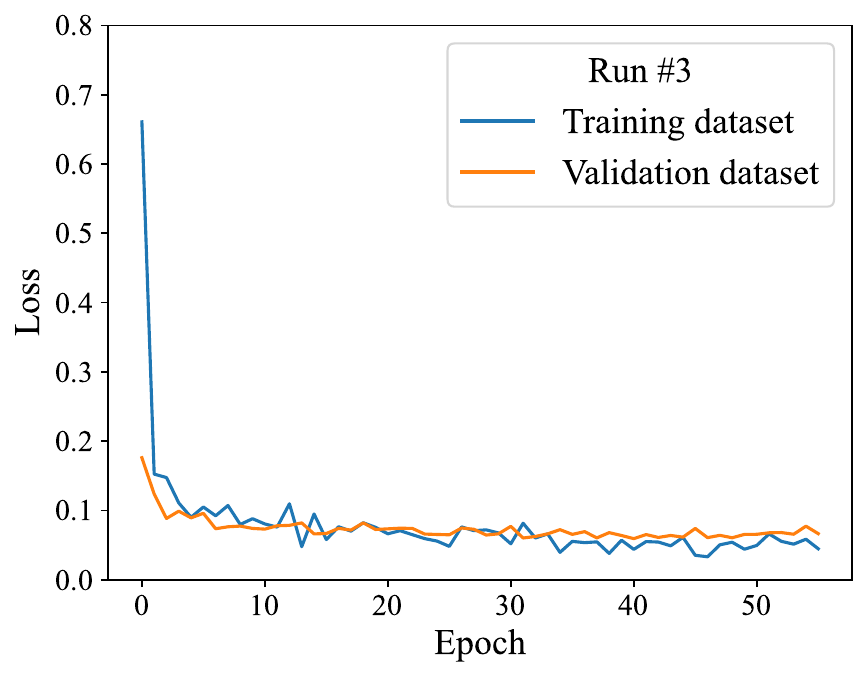}
    }\\
    \subfigure{
    \includegraphics[width=0.32\linewidth]{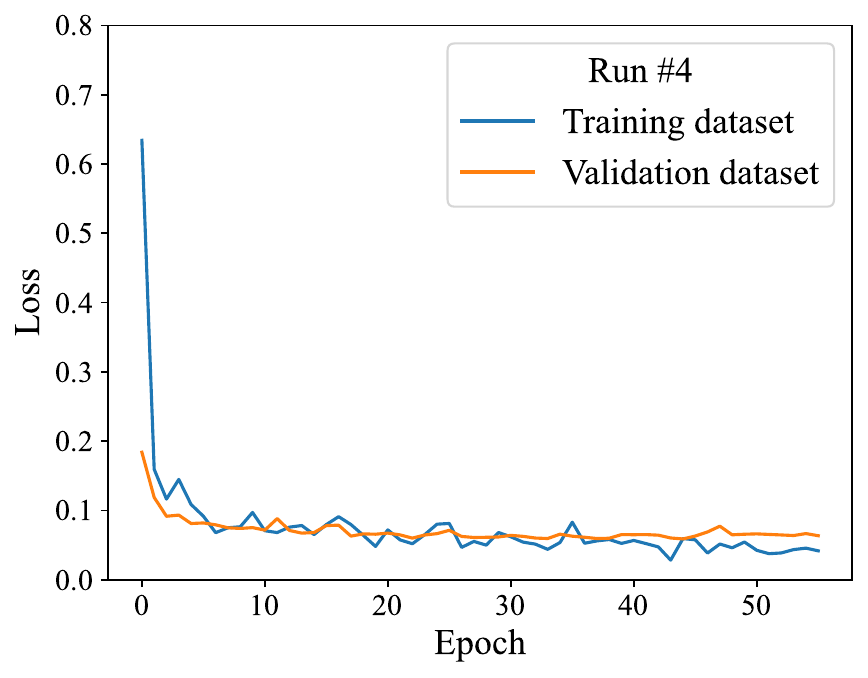}
    }
    \subfigure{
    \includegraphics[width=0.32\linewidth]{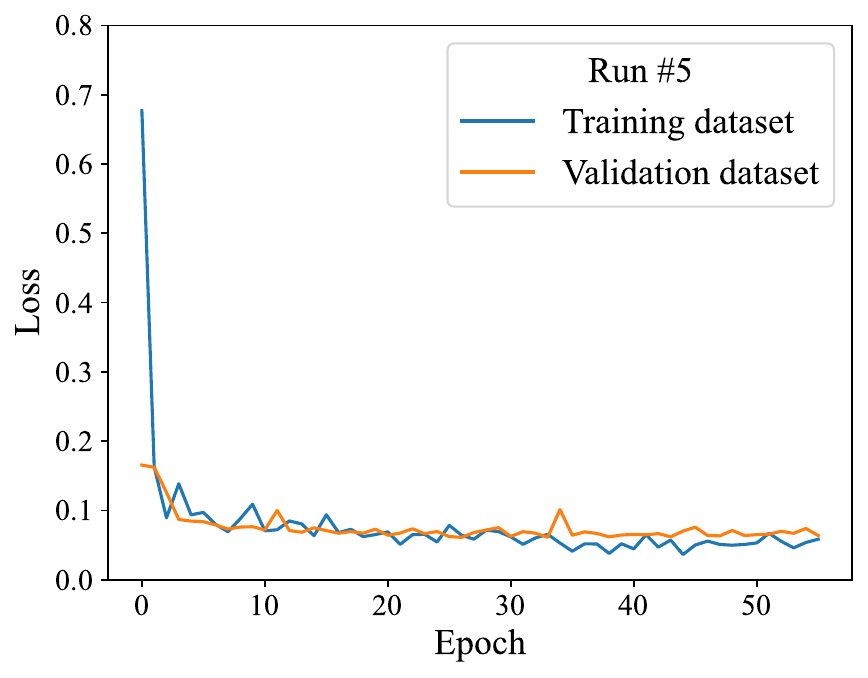}
    }
    \subfigure{
    \includegraphics[width=0.32\linewidth]{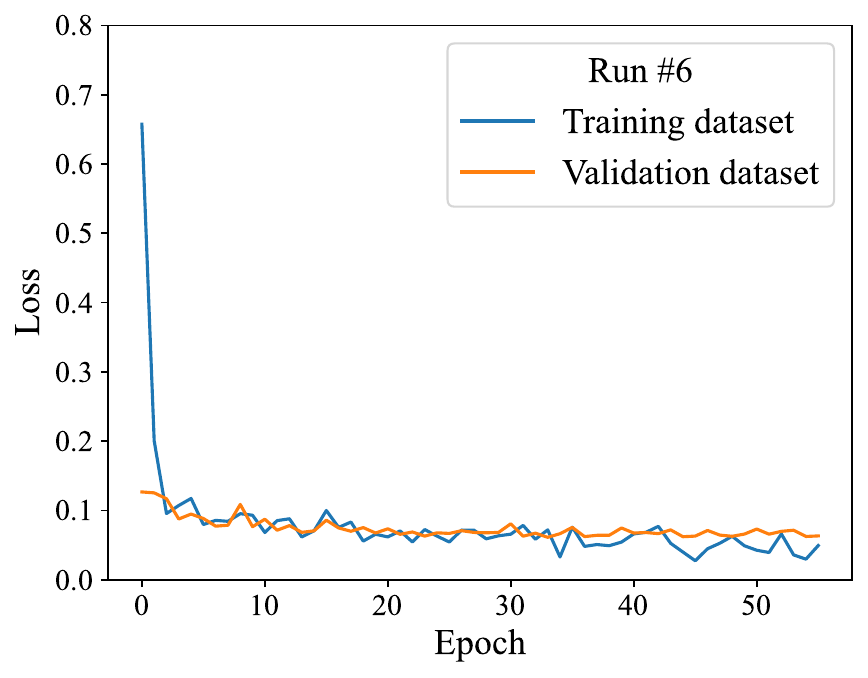}
    }\\
    \subfigure{
    \includegraphics[width=0.32\linewidth]{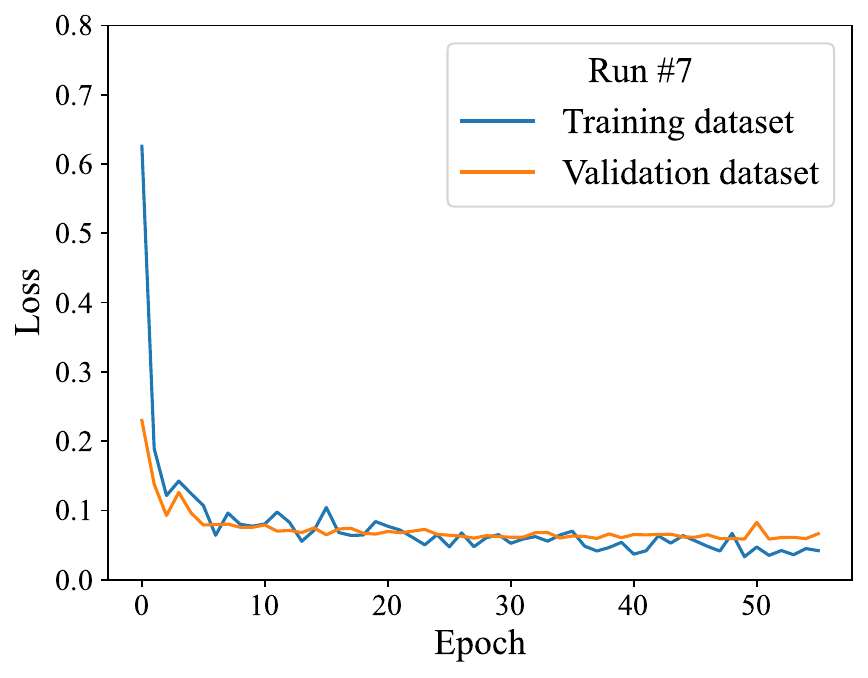}
    }
    \subfigure{
    \includegraphics[width=0.32\linewidth]{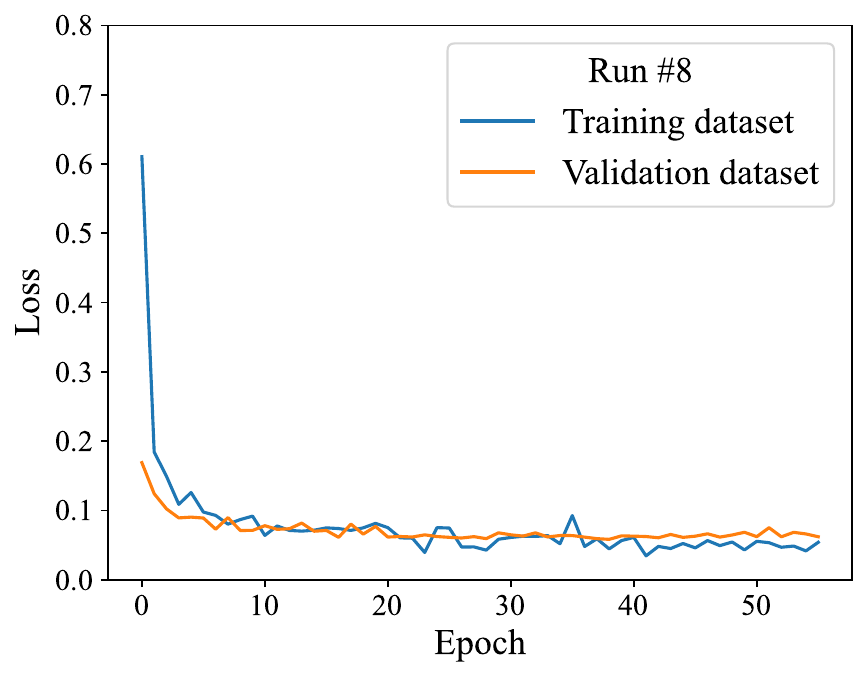}
    }
    \subfigure{
    \includegraphics[width=0.32\linewidth]{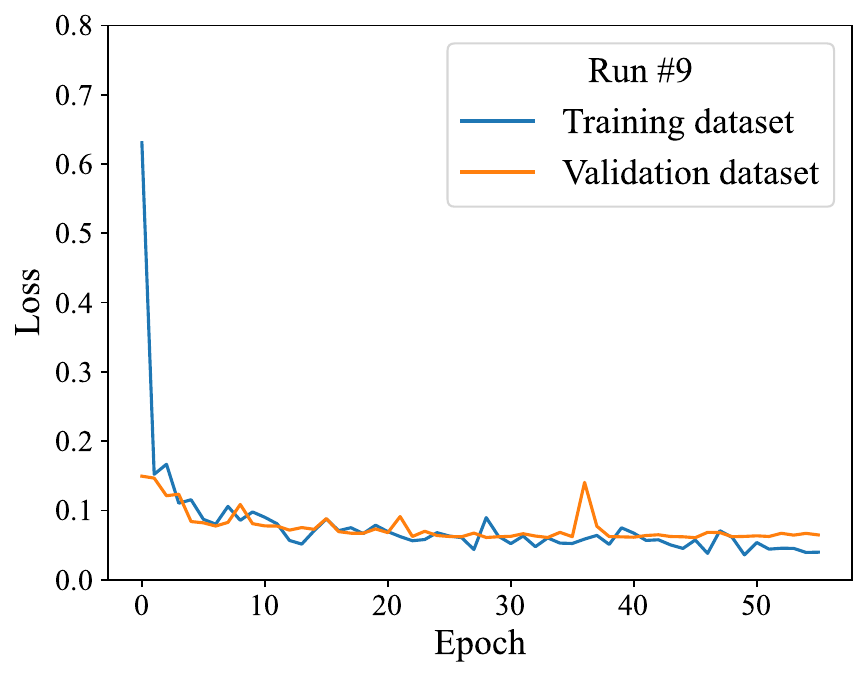}
    }\\
    \subfigure{
    \includegraphics[width=0.32\linewidth]{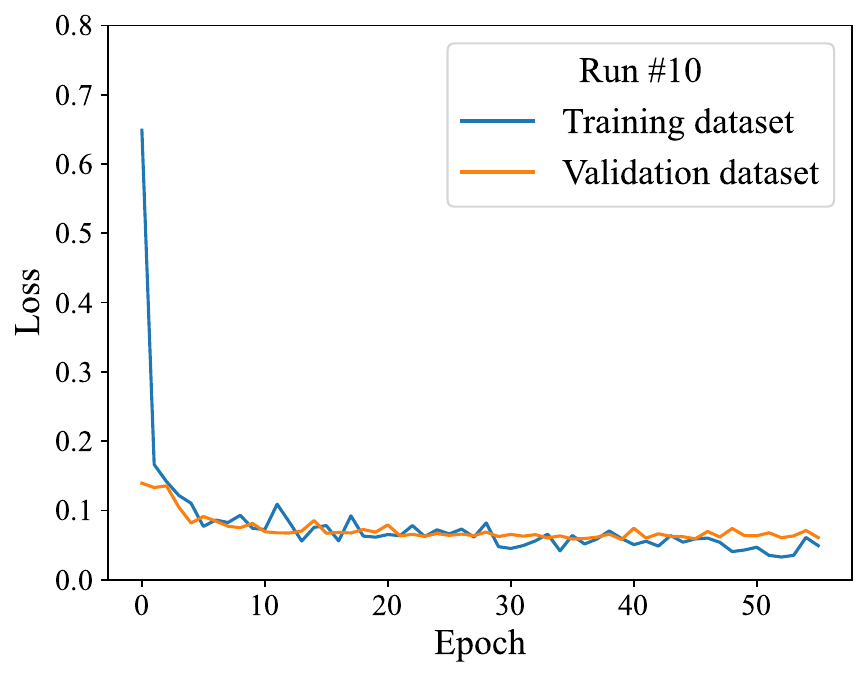}
    }
\caption{Results of the loss convergence test. For all ensemble learners, we see the loss for validation dataset follows the convergence of loss for the training dataset as desired, which shows no sign of overfitting or underfitting. \label{fig:losses}}
\end{figure*}

\begin{figure*}[t!]
\centering
\includegraphics[width=1.\linewidth]{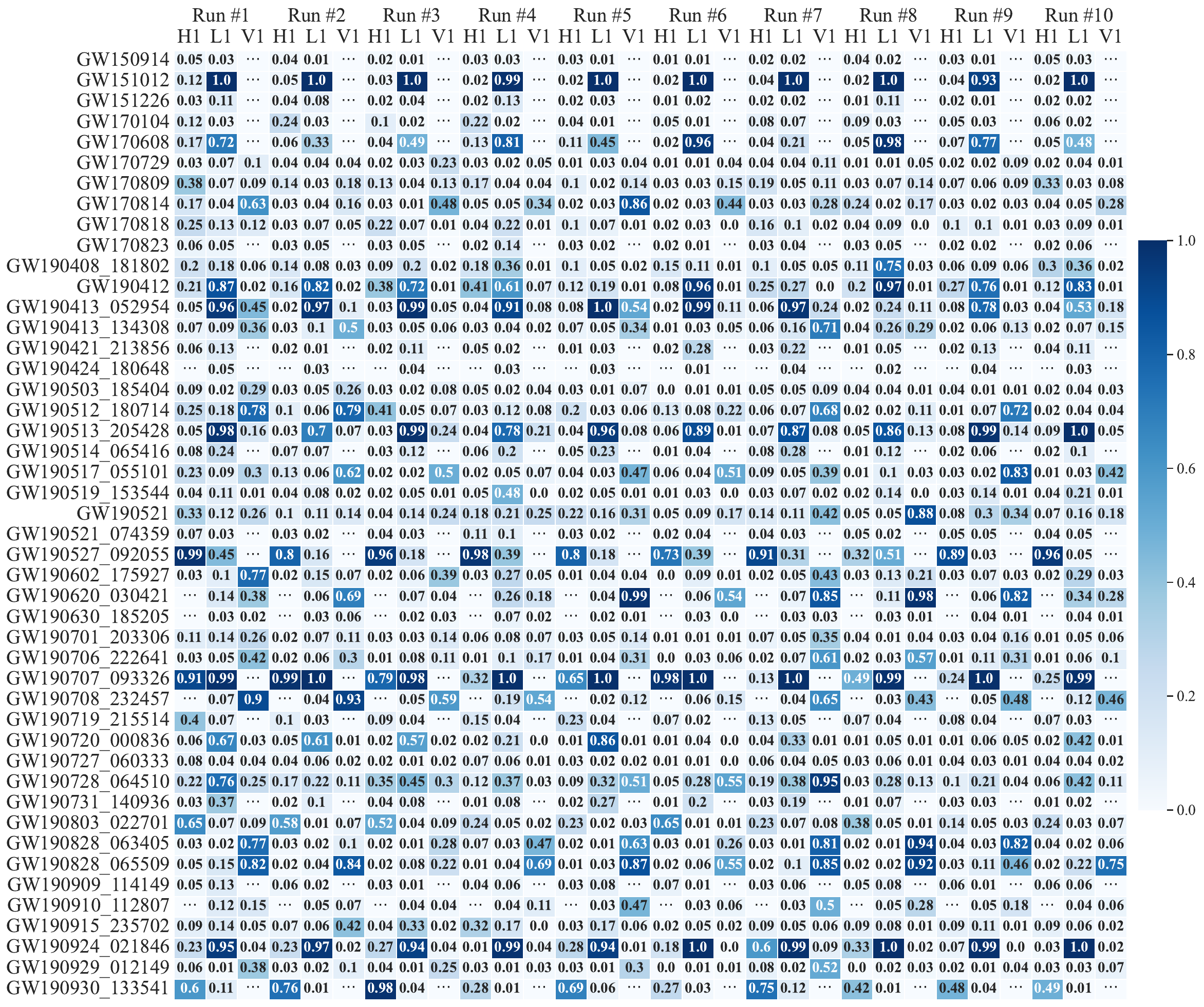}
\caption{Probabilities to the lensed class of all data evaluated from 10 ensemble learners, referred to ``Run \#''. We make the color of each cell to represent the annotated value of probability to the lensed class such as darker blue for values close to 1 or lighter blue for values close to 0. \label{fig:initial_probs}}
\end{figure*}

To check whether the trained model, i.e., each learner composing the ensemble learning, is either overfitted or underfitted, we examine the convergence of the loss over epochs for both training and validation datasets. As shown in Figure~\ref{fig:losses}, we can see that the loss for both datasets converges as the epoch evolves and the behavior of loss for the validation dataset is following the behavior for the training dataset. Therefore, we see no sign of overfitting or underfitting for all ensemble learners.

\section{Probabilities from Ensemble Learners}
\label{apx:result_initial}

We present the initial probabilities to the lensed class in Figure~\ref{fig:initial_probs} for all available data of the forty-six BBH events. We use the result for the determination of the initial classification tabulated in Table~\ref{tab:clf_result_ifo}.

\clearpage
\bibliographystyle{aasjournal}
\bibliography{astro,gw,method,lens}
\end{document}